# Spin-Reorientation Dynamics and Strong-Spin Phonon Coupling in Ce-substituted SmCrO$_3$

Shaona Das[1,3*†], Ravi Kiran Dokala[2,3*†] and Subhash Thota[3*]

[1]*Institute of Material Science, Technical University of Darmstadt, Darmstadt-64287, Germany*
[2]*Solid State Division, Department of Materials Science and Engineering, Uppsala University, Uppsala-75237, Sweden*
[3]*Department of Physics, Indian Institute of Technology Guwahati, Guwahati, Assam-781039, India*

## Abstract

We report the influence of Ce$^{3+}$ substitution on the magnetic structures and phonon dynamics in SmCrO$_3$ perovskites. Magnetic landscapes are spanned by long-range canted anti-ferromagnetism, AFM with Néel temperatures, $T_N$ ~ 196 K accompanied by spin-reorientation transitions, $T_{SRPT}$ at 42 K. In Sm$_{0.9}$Ce$_{0.1}$CrO$_3$ (SCCO), Ce$^{3+}$ substitution at Sm$^{3+}$ sites transform the weak ferromagnetic (FM) $\Gamma_4(G_x, A_y, F_z; F_z^R)$ state into robust AFM $\Gamma_1(A_x, G_y, C_z; C_z^R)$ configuration through a gradual crossover. Such coexistence of magnetic spin configurations ($\Gamma_1(AFM) \leftrightarrows \Gamma_4(WFM)$) results in the enhanced high coercive field and a pronounced exchange bias-field, $H_{EB}$ ~ 2 kOe. Spin-only driven magneto-crystalline anisotropy of Cr$^{3+}$ and spin-orbit driven magnetic moment in Sm$^{3+}$, and Ce$^{3+}$ exhibits spin-phonon coupling through $A_{1g}(6)$ mode in SCCO are consistent with the temperature dependent spectral features of the isostructural magnetic systems and quite significant in SCCO which is in accordance with the higher structural distortion in SCCO. These results demonstrate that site-specific $R^{3+}$ substitution modulates lattice distortions, spin–phonon coupling, and spin–orbit interactions, offering pathways to optimize perovskites for diverse spintronic applications.

[†]equal contribution

[*]Corresponding authors: Shaona Das, shaona.das@oxide.tu-darmstadt.de
Ravi Kiran Dokala, ravi.kiran.dokala@angstrom.uu.se
Subhash Thota, subhasht@iitg.ac.in

1.  **Introduction**

Heavier rare-earth (*R*) perovskites especially the *R*-chromates, $RCrO_3$ with *R* = Pr, Ce, Gd, Sm, Nd etc. exhibit complex magnetic ordering which holds robust spin dynamics and strong exchange interactions between $Cr^{3+}$-$Cr^{3+}$, $R^{3+}$-$Cr^{3+}$, $R^{3+}$-$R^{3+}$. These systems exhibit some fascinating physical phenomena such as negative magnetization, magnetization compensation, spin-flip, giant magneto-caloric effect and field-induced magnetic-phase transitions which play a vital role in designing spin-valve devices [1, 2]. Rare earth orthochromites have been examined dynamically over the past few decades using various diverse characterization techniques such as heat capacity, neutron diffraction, Mossbauer spectroscopy, magneto-dielectric and optical-absorption spectroscopy [3, 4, 5]. These investigations were widely utilised to probe the ferroelectricity driven by the co-existence of local non-central symmetry and multiple magnetic ordering (antiferromagnetic and ferromagnetic behaviour) in $RCrO_3$. Along with these properties, some other impeccable phenomena such as exchange bias, spin reorientation transitions, negative magnetization and its reversal, spin glass state etc., have been extensively studied in $RCrO_3$ related systems [4 - 8].

Multiple magnetic cations present in the perovskite drives the magnetic anomaly from conventional to exotic. Usually, the magnetic lattice structures are represented by Bertaut's Gamma notations in perovskites, among which $\Gamma_1(A_x, G_y, C_z), \Gamma_2(F_x, C_y, G_z)$ and $\Gamma_4(G_x, A_y, F_z)$ are well known [9]. Interestingly, the $\Gamma_2$, $\Gamma_4$ consists of weak FM alignment in one of their axes and $\Gamma_1$ doesn't contain FM alignment in any of the axes. The behaviour has been well established through the field-dependent magnetization results carrying a well-defined coercive fields and remanence values. From the previous reports, among the family of orthochromites, $Y_{1-x}Pr_xCrO_3$ exhibits purely AFM spin structure with no remanence of FM axes which is identified as $\Gamma_1$ spin structures [10]. This highly implies the possibility of such magnetic spin structures further in the family of orthochromites. Interestingly, the field-manipulated magnetic structures carrying a distinguished magnetic ordering like purely AFM $\Gamma_1$ and $\Gamma_4$ with AFM and weak FM moments lead to the phenomenon of exchange bias. Therefore, *A*-site manipulated $SmCrO_3$ caught our attention due to its' multiple competing magnetic lattice structures from the previous literatures [11].

$SmCrO_3$ and $CeCrO_3$ are very well-known antiferromagnetic systems exhibiting the magnetization spin reversal behaviour [3, 4]. Previous reports show $SmCrO_3$ aligns *G*-type AFM with its $T_N$ ~ 191 K and $T_{SRPT}$ ~ 34 K [12] and $CeCrO_3$ on the other hand exhibits the *G*-type AFM behavior with the spin-flip transition gradually occurring at higher magnetic fields and compensating for the magnetization and ventures into the positive scale. Cao et. al. reports the $T_N$ ~ 230 K and $T_{Comp}$ ~ 100 K with spin flip transitions occurring at $T_{SF}$ ~ 36 K at 1.2 kOe [3].

In this article, we report an investigation of the polycrystalline chromate perovskite SCCO through their structural, electronic, magnetic properties. X-ray diffraction and Raman studies contribute to the understanding of the internal atomic atmosphere and their response to optical spectroscopic techniques. They exhibit different magnetic spin-orientations sensitive to externally applied temperature and magnetic field, and exchange bias with multiple magnetic landscapes. Path independent spin-flip transitions are studied with conventional magnetization protocol like (cooling and heating in unequal fields (CHUF). Further, one of the investigated systems displays strong spin-phonon coupling due to the local octahedral distortions mainly caused by the substitution of trivalent Ce at the Sm sites which are having relatively different in size.

**2.     Experimental details**

Polycrystalline SCCO was synthesized by standard solid-state reaction method. The stoichiometric proportions of powder precursors of $CeO_2$ (99.95%), $Sm_2O_3$ (99.99%) and $Cr_2O_3$ (99.99%) were ground in an agate mortar with pestle in air for 5 hours, calcined at 1000 °C for 24 hours under the ambient conditions in order to prepare the highly homogeneous mixture of SCCO. After the calcination the sample was then ground again for 2 hours to break the probable bonding formed due to calcination and was pelletized and sintered at 1200°C for 24 hours in air using a tubular furnace from Nabertherm GmbH (Germany) followed by natural cooling to room temperature. Phase purity of the final sintered samples was confirmed with the help of room temperature Rigaku; TTRAX III XRD equipped with 1.54056 Å wavelength laser source of Cu-$K_{\alpha 1}$ radiation with step size 0.02. In order to evaluate the charge states of the core level constituent elements, XPS experiment was performed using the high-resolution PHI 5000VVersaProbe III Scanning XPS with microprobe Al $K_\alpha$ (544 eV). DC magnetization measurements were performed using a Dyna-Cool model physical property measurement system (PPMS) from Quantum designs. The magnetization values were recorded at different external fields (50 Oe to 20 kOe) with respect to temperature and at variable temperatures (3 K to 300 K) with respect to field under different protocols such as zero-field cooled warming (ZFCW), field cooled cooling (FCC), field cooled warming (FCW). Isothermal field dependent magnetization hysteresis loops were recorded in the range of +90 kOe to -90 kOe at selected temperatures. The room temperature Raman spectra was recorded within the range of 100 cm$^{-1}$ to 800 cm$^{-1}$ and separately 100 cm$^{-1}$ and 200 cm$^{-1}$ by means of HR Raman spectrometer from Horiba Jobin Yvon (model: LabRam HR) with the excitation of He-Ne laser of fixed power 20 mW. The THMS600 module was supplemented for the temperature variant readings up to 80 K for both the samples within the lower range of wavelength.

**3.     Results and Discussion**

**3.1.    Crystal structure analysis**

The following complex oxide perovskites such as $CeCrO_3$ and $SmCrO_3$ are generally stable in slightly distorted orthorhombic perovskite structure with space group *Pbnm* [13]. Substituting the $SmCrO_3$ with larger size Ce adjusts itself at the *A*-site for both the pristine compounds in place of Sm with 12-fold coordination while *B*-site remains unaltered with Cr with 6-fold coordination. Figure 1 displays the XRD pattern for SCCO which confirms the phase purity without any signature of secondary impurity phase. The XRD patterns are analysed with the help of Rietveld refinement technique using FULLPROF software [14]. The Bragg peaks are indexed according to the *Pbnm* (no. 62) space group and the obtained goodness of the fitting $\chi^2 = 2.63$ indicates the reliability of the fitting. The occupation number of (Sm, Ce) are refined during the fitting in order to justify the *A*-site cation ratio and the refined value of Sm/Ce are nearly same stoichiometric proportions as that of the theoretical compositions considered initially. We can observe the trend of fluctuated increase in the lattice parameters *a*, *b* with the substitution of higher ionic radii cation Ce at Sm site. Consequently, the volume decreases by 0.39 % for Ce doping at Sm sites due to the effective iodic radius of $Ce^{3+}$ (1.143 Å) being larger than the $Sm^{3+}$ (1.079 Å) with 8-coordinatio number while the in-plane parameter, $c$ increases. In order to understand the internal crystallographic behaviour, we tabulated all the structural parameters refined and calculated along with the comparative study of the pristine compounds from the previously reported values in Table 1. These changes can be explained with the help of difference of average radius of $A$-site cation: $r_{avg} = \sqrt{[(0.9) \times R^2_{Sm^{3+}}] + (0.1 \times R^2_{Ce^{3+}})}$ for SCCO which increases to 1.086 Å compared to 1.079 Å for pristine compound with $R_{Sm^{3+}} = 1.079$ Å having VIII coordination number and that of $R_{Ce^{3+}} = 1.143$ Å having the same coordination number [15]. The substitution of larger cation impacts the crystal symmetry by introducing significant distortion of the unit-cell which is popularly known as the factor of tolerance ($t$), given in the following Eq. (1).

$$t = \frac{\{[(1-x) \times R_{Sm^{3+}}] + (x \times R_{Ce^{3+}}) + R_{O^{2-}}\}}{\sqrt{(R^{3+}_{Cr} + R^{2-}_O)}} \qquad 1$$

Accordingly, the octahedral distortion (Δ) can be expressed as below in Eq. (2).

$$\Delta = \left(\frac{1}{N}\right) \sum_{n=1,N} \{(d_n - \{d\}) \sphericalangle d >\}^2 \qquad 2$$

Here, the apical Cr-O$_{(1)}$-Cr bond angle ($\theta_1$), basal Cr-O$_{(2)}$-Cr bond angle ($\theta_2$) and the tilt angles θ and ϕ inside the CrO$_6$ octahedra along the pseudo-cubic axes [110] and [001] calculated from the bond angles that exhibit visible deviations with the $A$-site cation substitution than that of the average bond lengths between the Cr and the O anions directing towards the bond rigidity expected in octahedral symmetry surrounding the trivalent Cr [22]. Also, we have calculated the tilt angles from the experimentally obtained lattice parameters, $\theta = cos^{-1}\frac{\sqrt{2}a}{b}$, $\varphi = cos^{-1}\frac{a}{b}$ and the magnitudes match well with the previously reported values of similar systems [16, 17]. Along with the tilting of the CrO$_6$ octahedra, due to the mismatch between the A-site Sm/Ce and B-site Cr cations, an impulsive reduction in the strain parameter $s$ with an increment of ~ 2.2 % is observed as the average radius of the $A$-site cation comprises the Wyckoff position eccentricities of Sm/Ce (4c) and the Cr (4b) sites.

### 3.2. Electronic Properties

In order to probe the electronic structure of the synthesized perovskites, we performed XPS studies through which we analysed the composition of the SCCO with exact electronic states possessed by the constituent elements and validate the phase purity. Figure 2 (*i – iv*) demonstrates the room temperature XPS spectra, where the scattered symbols are the experimental data, and the curves are the fitted lines with the combination of Lorentzian and Gaussian functions. The collected high resolution XPS spectra possess the binding energy resolution of 0.1 eV. We used the tougaard algorithm for the background correction and a nonlinear least square fitting procedure for resolving all the constituent chemicals. The core level of the spectra is calibrated with the C-1$s$ binding energy 285 eV. We examined all the surface level particle electronic configuration of the constituent elements in SCCO. The binding energy dependence of the ejected photoelectron intensity is probed for each of the elements separately: Sm-3$d$, Ce-3$d$, Cr-2$p$ and O-1$s$. All the binding energies magnitudes are persistent with the standard values of National Institute of Standards and Technology (NIST) database [18]. From Fig. 2(*i*) the Sm-3*d* core level spectrum is deconvoluted into two peaks centred at 1080.7 eV and 1108.4 eV. Among these, the lower binding energy peak is associated with the $3d_{5/2}$ and the peak at higher energy is originating due to $3d_{3/2}$ state. These two peaks are separated by the large binding energy Δ ~ 27.5 eV is generally characterized as the spin-orbit coupling which confirms the trivalent oxidation state of Sm [19]. Figure 2(*ii*) shows the XPS spectra of Ce-3*d* core-level spectrum with deconvoluted peaks at 881.4 eV, 885.2 eV, 899.2 eV and 903.3 eV. The core-level consists of two doublets, $3d_{5/2}$ and $3d_{3/2}$ electronic states, and we identified them based on the literature available. Ce possess the Xenon like configuration with filled *s* orbital and singly occupied *f* and *d* orbitals that are easily prone to get oxidised to $Ce^{4+}$ resulting in the signature of both the states in the XPS spectrum [20]. But in our experimental data, we did not find any signature of $Ce^{4+}$ confirming the phase purity. The peaks positioned at 881.4 eV and 885.2 eV signifies the spin-orbit coupling induced $3d_{3/2}$ state and the peaks at 899.2 eV and 903.3 eV specifies the $3d_{5/2}$ core level [20]. On the other hand, the Cr-2*p* spectra given in Fig. 2(*iii*) consists of 4 peaks at 576.2 eV and 577.8 eV, 586.1 eV and 586.9 eV which are consistent with the range given in the standard XPS database of NIST [21]. The peaks at 586.1 and 586.9 eV are originating due to the doublet of $2p_{1/2}$ electronic state and the other two peaks are ascribed to the $2p_{3/2}$ state with no signatures of any satellite peaks. In this case, the spin-orbit splitting is ~ 10.2 eV and 10.4 eV validating the existence of the trivalent state of Cr ion. The mere difference with the NIST database standard peak positions can be due to the change in $r_{avg}$ resulting in the modified bond length in the $CrO_6$ octahedra. Finally, the asymmetrical curve of O-1*s* core-level spectra raised two peaks and another extra hump originated as a result of surface absorbed oxygen species. The spectrum is well fitted with 23% Lorentzian and 77 % Gaussian fitting with $\chi^2$ value 2.01. The FWHM value was kept fixed at 1.03 for the best fitting purpose. The fitted peaks are indexed at 529.5 eV, 530.1 eV and 531.9 eV. The spectrum at 529.5 eV is the result of saturated $O^{2-}$ ions along whereas the peak at 530.1 eV confirms the O1*s* state [22]. The remaining one locating around 531.9 eV indicates towards the surface originated excess oxygen [23]. This is understood that the peak values are little higher as compared to the ideal cases [24, 25, 26]. We did not observe any additional transitions pertaining to the electronic state of constituent elements studied in this case which confirms the purity of the sample.

### 3.3. Magnetic Structure

#### A. Temperature dependent magnetization spin-reversal and spin-reorientation

Figure 4(a) represents the temperature dependent magnetization of SCCO under ZFCW at an applied field of 100 Oe and 200 Oe. It exhibits the long-range magnetic ordering that suggests the canted anti-ferromagnetism with Néel temperature at $T_N$ ~ 196 K which is confirmed by the $\frac{d(\chi T)}{dT}$ vs T with Curie-Weiss linear fit that ventures into the negative temperature scale and meets the *X*-axis at Curie-Weiss temperature, $\theta_D$ ~ -663 K by extrapolating the linear fit at the high temperature paramagnetic behaviour [12]. Long-range magnetic ordering lead by the $Cr^{3+}$ sublattice configured in the G-type CAFM structure with $\Gamma'_4$ $(G_x, A_y, F_z)$. $Sm^{3+}$ and $Ce^{3+}$ moments align anti-parallel to the local field created by the $Cr^{3+}$ sublattice. The beginning of the reduction of magnetization values from peak point to the lower temperatures and lower magnetization values is beginning of the $R^{3+}$ moments alignment. As the $R^{3+}$ readily aligns, the mutual antiparallel coupling between $R^{3+}$-$Cr^{3+}$ is undergoing a spin-reorientation phase transition into $\Gamma_1(A_x, G_y, C_z; C_z^R)$ at $T_{SRPT} \sim 41.9$ K [12]. This active transition is identified with the shaded region during the slope change in MT plot. The spin-reorientations with similar $T_{SRPT}$ values and similar slope change regions highlighted with shaded background which is evidently investigated with different magnetic fields of 200 Oe, 500 Oe, 1500 Oe both in the FCC and FCW protocols as shown in Figure 7(b). The overall magnetization of the $Cr^{3+}$ and $R^{3+}$ sublattices increases with the applied magnetic field values which is conventional in these complex chromite oxides.

## B. Spin-reorientation and Exchange-bias

Figure 5(a) and 5(b) shows measurements of SCCO under CHUF protocol. During the FCC protocol under the application of 10 kOe, the $Cr^{3+}$ begins to align into the $\Gamma_4(G_x, A_y, F_z)$ spin structure whose resulting magnetisation is parallel to the applied field. Towards reasonably low temperature, the rare-earth component, $\Gamma_4(F_z^R)$ begins to align. As the $Cr^{3+}$ is AFM dominant lattice, it contains an AFM axis which creates a local field to the $Sm^{3+}$/$Ce^{3+}$ sublattice [27, 28]. The $R^{3+}$ sublattice aligns antiparallel to the $Cr^{3+}$-sublattice which brought a gradual decline in the overall magnetization from the observed peak magnetization to temperatures below the $T_{SRPT}$. Here, even after reaching the low temperature 3 K, it doesn't show such sharp reduction in the magnetization value as seen in the FCC of $H_{DC} = 200$ Oe case. Such behaviour is obvious as the nucleation of the $\Gamma_1(A_x, G_y, C_z; C_z^R)$ spin structure is hindered by the strengthened ferromagnetic components in the $\Gamma_4(G_x, A_y, F_z; F_z^R)$ spin structure due to the field, 10 kOe. More number of frozen $\Gamma_4$ structures persist in the low energy $\Gamma_1$ phase below the $T_{SRPT}$. Now, at 3 K, the field is reduced to as low as 500 Oe and the magnetization is recorded under FCW protocol. The observed magnetization behaviour exhibits a completely transformed $\Gamma_4(G_x, A_y, F_z; F_z^R)$ into $\Gamma_1(A_x, G_y, C_z; C_z^R)$ phase, and the magnetization is quite similar to the FCW magnetization at 500 Oe. This also signifies that the FCW at 500 Oe doesn't get affected with the FCC measurement at 10 kOe field.

In contrast, from Fig. 5(b) During the FCC protocol, under the application of 70 kOe, the magnetization is quite similar till $T_{SRPT}$, but below $T_{SRPT}$, the magnetization increases to higher values. One of the two possible reasons can be, the frozen FM configuration of $\Gamma_4(F_z; F_z^R)$ phase ceases to align into AFM configuration of $\Gamma_1(C_z; C_z^R)$. Secondly, the $Sm^{3+}$/$Ce^{3+}$ sublattice which began to align in the antiparallel setup couldn't overcome the field which is as high as 70 kOe and instigates a parallel alignment *w.r.t.* the applied $H_{DC}$ direction. Overall, the $\Gamma_4(G_x, A_y)$ may transform into $\Gamma_1(A_x, G_y)$ but the $\Gamma_4(F_z; F_z^R)$ doesn't allow $\Gamma_1(C_z; C_z^R)$ [9]. Besides these incomplete transformations of the gamma structures, the higher magnetic field freezes more percentage of $\Gamma_4$ even into the $\Gamma_1$ region more impactfully. Again, at 3 K when the $H_{DC}$ is reduced to as low as 500 Oe, the $\Gamma_1(A_x, G_y, C_z; C_z^R)$ phase is completely achieved and the magnetization interestingly represents the unaffected behaviour of FCW due to the treatment of fields as high as 70 kOe during FCC.

Figure 5(c) represents magnetization under constant field $H_{DC} = 500$ Oe and different temperature sweeps. During the first cycle, under FCC, the magnetization recorded from 300 K to till 15 K, which is very well below the $T_{SRPT}$. This results in relatively more conversion of $\Gamma_4 \rightarrow \Gamma_1$ spin structures. But the conversion is not complete, and during the back sweep from 15 K to 300 K, under FCW protocol, from 15 K to $T_{SRPT}$, more $\Gamma_4$ will be converted to $\Gamma_1$ till $T_{SRPT}$. As the temperature crosses the $T_{SRPT}$, $\Gamma_1 \rightarrow \Gamma_4$ conversion initiates and rate of conversion is slow when compared to FCC protocol which signifies a first-order phase transition and substantial reduction in the magnetization is observed in FCW compared to FCC. Now, from 300 K, during the second cycle of measurement, the temperature is swept till 40 K which is the closest neighbourhood of $T_{SRPT}$. This FCC protocol, the converts very few $\Gamma_4$ into $\Gamma_1$. As this conversion will be less when compared to the first cycle, the formed loop is narrow but still exhibits a first-order phase transition. Interestingly, during the third cycle, under FCC, the temperature is swept till 90 K from 300 K, and the magnetization is measured in back sweep till 300 K. Here as the $T_{SRPT}$ is reasonably far from 90 K, the $\Gamma_4$ persists and doesn't exhibit any first-order phase transition all through the FCC and FCW protocols. This kind of behaviour is reported as "magnetic glass" where FM/$\Gamma_4$ structures freeze into AFM/$\Gamma_2$/$\Gamma_1$ regions in $Gd_{0.5}Sm_{0.5}CrO_3$, $La_{0.5}Ca_{0.5}MnO_3$, $Pr_{0.45-x}Yb_xSr_{0.55}MnO_3$, etc [29, 30, 11]. The spin-reorientation transition. $T_{SRPT}$ is the super-cooling limit, below which $\Gamma_4$ phase which is considered as a liquid phase that crystallizes into $\Gamma_1$ phase. In such systems, investigation carried out through CHUF measurements gives highly productive results to confirm the glassy signatures of these magnetic islands. In the following figure, we will find out the impact of the spin-reorientation dynamics on the field-dependent magnetization measurements.

Field-dependent magnetization measurements were performed under ZFC protocol, FC protocol of 70 kOe, and FC protocol of -70 kOe at temperature 25 K as shown in the Fig. 5(d). During the ZFC case, the SCCO exhibits a positive exchange bias field, $H_{EB}$ ~ +945 Oe. During the FC at +70 kOe, $H_{EB}$ ~ +2190 Oe, and at -70 kOe, $H_{EB}$ ~ -2352 Oe. The exchange bias during the ZFC case reflects inherent spin-dynamics of $\Gamma_4$ and $\Gamma_1$ phases. Under the ZFC protocol at 25 K, during the first cycle of zero Oe $\rightarrow$ 90 kOe, at the lower fields, magnetization shows up corresponding to the $\Gamma_1$ phase which is completely transformed into $\Gamma_4$ phase at the higher fields. While tracing back during the second cycle, from 90 kOe to 0 kOe, the $\Gamma_4$ phase persists till 0 kOe which stands as a reason for the remanence, $M_R$ ~ 0.75 $\frac{\mu_B}{f.u}$ exhibited in this MH-curve. The remanence is not only a contribution from $\Gamma_4$ phase but also with the $\Gamma_1$. The remanence would've been more if it is contributed by $\Gamma_4$ phase alone. The contribution from $\Gamma_1$ phase actually brings the magnetization down more quickly as the building blocks of $\Gamma_1$ phase is purely AFM. The same trend continues and makes the magnetization zero with the negative magnetic-field giving a coercive-field, $H_C$ ~ -9.54 kOe. During the third cycle, 0 Oe $\rightarrow$ -90 kOe, the magnetization takes a quicker response towards the saturation magnetization which makes the $Sm_{0.9}Ce_{-0.1}CrO_3$ attain a higher magnetization value at -90 kOe [31]. During the fourth cycle, from -90 kOe $\rightarrow$ 0 kOe, results in showing higher remanence, $M_R$ ~ 1.089 $\frac{\mu_B}{f.u}$, and which further influence the coercive field to be shifted to higher fields, $H_C$ ~ 14.01 kOe. These spin-reorientation dynamics between the $\Gamma_4$ phase and $\Gamma_1$ phase exhibits a shift in the *MH*-loop towards the positive field axis, resulting in a positive exchange bias field value, $H_{EB}$ ~ 945 Oe. In order confirm such rare positive exchange bias, we performed *MH* at FC protocol with -70 kOe and +70 kOe, which give positive exchange bias field and negative exchange bias field respectively and confirms the characteristic exchange bias in the SCCO.

### C. Spin-reorientation from WFM-$\Gamma_4 \rightarrow$ Purely AFM-$\Gamma_1$: role of magnetic field

Figure 6 represents the field dependant magnetization in SCCO at different temperatures under ZFC protocol. The *M-H* curve at 3 K exhibits a purely-AFM phase in the form of $\Gamma_1(A_x, G_y, C_z)$ spin structure which aligns with the $Cr^{3+}$-sublattice whose magnetic axis is parallel to the applied magnetic field. Interestingly, the magnetization achieved at the 90 kOe is 12.5 $\frac{\mu_B}{f.u}$ which is higher than any other magnetization achieved at other temperatures. The conventional alignment of $R^{3+}$-sublattice is anti-parallel to the local field created by the $Cr^{3+}$-AFM sublattice. Contrastingly, the behaviour the $R^{3+}$-sublattice takes the parallel configuration with $Cr^{3+}$-local field and externally applied magnetic field such that a continuous increase in the overall magnetization is reinforced into the *M-H* curve. Similar increase in overall magnetization is hindered by the incomplete formation of $\Gamma_1$ spin-structures, which further triggers field-dependent transition from $\Gamma_1 \rightarrow \Gamma_4$ at higher fields that is noticed with a significant tilt in *M-H* curve raising the magnetization slightly from the linear behaviour. As the percentage of formation of $\Gamma_4$ is very less with 90 kOe in the low temperature $\Gamma_1$ regime, the magnetization seems reversible and shows almost the similar behaviour like *MH* at 3 K. Substantial percentage of $\Gamma_4$ spin-structures exists at 25 K, which leads to a WFM-$\Gamma_4$ behaviour with remanence and showing intriguing values of ZFC and FC exchange bias as discussed from the Figure. 6(d). As the temperature progresses the $\Gamma_4$ phase dominates and removes the signatures of $\Gamma_1$ from the MH-curve. Purely-AFM configuration of $\Gamma_1(A_x, G_y, C_z)$, consisting *A*-type AFM in *x*-direction, *G*-type AFM in *y*-direction, and *C*-type AFM in *z*-direction dominates at 3 K as it is far below the $T_{SRPT}$ which creates a suitable environment for the lowest energy state for $\Gamma_1$ phase to exist. Followed by a mixture consists more of $\Gamma_1$ and few $\Gamma_4$ spin-structures to coexist and exhibit a field-induced transition at 10 K. Near to $T_{SRPT}$, at 25 K gives an interesting and desirable ZFC-exchange bias contributed by the dynamics of $\Gamma_1$ and $\Gamma_4$ phases [3]. From Fig. 6(*i*), the application of magnetic field, 70 kOe after FCC in 500 Oe and bringing it back to the 500 Oe to proceed further with FCW doesn't bring any change in the $\Gamma_1$ phase showing it's reversible behavior.

### 3.4. Temperature dependent Raman scattering

Raman spectroscopy is a highly sensitive tool to probe the atomic-vibrations and -rotations, and additional low-frequency aspects in orthorhombic perovskite oxides, offering critical insight into octahedral tilts, cation displacements, and bond strength variations that govern their electronic and magnetic properties. This technique also elucidates the local crystal distortions and anomalous change in the local atomic environment in the compound due to its high sensitivity towards crystallographic symmetry change. Hence, an attempt has been made here to probe the modifications of the structural distortions that can lead to the measurable Raman shift due to the two different atoms of different cationic size, $Ce^{3+}$(1.143 Å) and $Sm^{3+}$(1.079 Å). In rare-earth chromite perovskites, where the *B*-site is occupied by $Cr^{3+}$ ions, Raman-active phonons are directly linked to the stability of the $CrO_6$ network and its coupling with the A-site cation environment.

Figure 7(a) displays the temperature dependent Raman spectroscopy of SCCO recorded between 80 K and 223 K, within the wavenumber range 100 cm$^{-1}$ to 800 cm$^{-1}$. We have followed the orthorhombic *Pbnm* structure following the Glazer's tilt system $a^-b^+a^-$ in which octahedral rotations occur about the crystallographic axes with mirror symmetry along the [010] plane [15]. The theoretical group analysis predicts decomposition of normal modes of *Pbnm* crystal structure to be 60 in number where the active 24 Raman modes are: $7A_{1g} + 5B_{1g} + 7B_{2g} + 5B_{3g}$. These 24 Raman active modes are categorised in two symmetric ($A_g + B_{1g}$), four antisymmetric octahedral stretching modes ($2B_{2g} + 2B_{3g}$), four bending modes ($A_g + 2B_{2g} + B_{3g}$), six rotation or tilt modes of the octahedral ($2A_g + B_{1g} + 2B_{2g} + B_{3g}$) and eight modes associated with well-defined *A*-site cation movement ($3A_g + 3B_{1g} + B_{2g} + B_{3g}$). The remaining modes consist of 25 infrared-active, 8 silent and 3 acoustic modes. The active modes typically originate from three structural perturbations:1) distortion of the CrO$_6$ octahedra 2) Jahn-Teller (J-T) distortion 3) Displacement of the *A*-site cation [32, 16]. In contrast to manganites or cobaltites, where mixed valence or orbital degeneracy favours J–T effects, chromites with Cr$^{3+}$ ([Ar]3d$^3$) exhibit no J–T activity due to their half-filled $t_2$g manifold and empty $e_g$ orbitals, unlike Cr$^{2+}$ (d$^4$) systems that undergo strong J–T splitting in both high- and low-spin states. Consequently, the rigidity of the Cr–O framework in *R*CrO$_3$ can be attributed primarily to octahedral tilt and A-site driven distortions, reinforcing the structural stability of these systems. [32]. Within the experimental range, 11 distinct Raman-active modes are clearly observed, showing systematic red- and blue-shifts with temperature in both samples, reflecting distinguishable anharmonic phonon interactions.

Low-frequency phonons (< 200 cm$^{-1}$) are dominated by rare-earth *A*-site cation displacements along *x* and *z* direction relating to the reduced mass (μ) of the *A*-site cation as: ω = √(k/μ) with, k as the effective force constant. The two modes under this category are labelled as $A_{1g}(1)$ and $B_{2g}(1)$. Above 200 cm$^{-1}$ the observed phonon modes correspond to vibration of Sm and O atoms, and bending and stretching of Cr-O bonds. Notably, $A_{1g}(2)$ and $B_{2g}(2)$ doublet modes reflect collective octahedral rotations around the *y*-axis, a key feature of the $a^-b^+a^-$ tilt system. Among the higher-frequency modes, the most prominent and intense mode at ~ 561 cm$^{-1}$ (80 K) is related to the octahedral bending inside the unit lattice cell whereas the second most strengthened mode at 698 cm$^{-1}$ (at 80 K) corresponds to the antisymmetric stretching vibrations of CrO$_6$ octahedra, a mode highly sensitive to Cr–O bond length and bond-angle distortions. The persistence of these modes and their moderate temperature evolution indicate the absence of structural phase transitions within the measured window, reinforcing the stability of the orthorhombic phase [33].

Among the observed modes, we explain $A_{1g}(6)$ mode shifting towards the lower wavenumber with increasing the temperature towards room temperature indicates towards the Sm(Ce)/Cr-O bond lengths and increase in bond-angle O-Cr-O. Figure 7(b) and 7(c) represent the zoomed version of fitted $A_{1g}(6)$ mode with their fitted line parameters around $T_N$ defined as peak position in cm$^{-1}$ with anharmonic equation as a function of temperature . To further scrutinize the Raman spectra for existence of correlation between vibrational and magnetic degrees of freedom which is termed as spin-phonon coupling, we have fitted the Raman spectra with Lorentzian function to extract the temperature dependence of line shape parameters for highly raman active $A_{1g}(6)$ mode within the investigated temperature region. As shown in Fig. 7, a clear abrupt deviation of ω(*T*) from the anharmonic baseline at $T_N$ (panel d) below $T_N$ the mode hardens relative to the purely anharmonic expectation indicates to the spin-phonon anomaly signifying strong spin-phonon coupling. Comparable observations in $RCrO_3$ have been reported and attributed to the role of magnetic $R$-ions and structural details. We also observed the broadening of the modes while reaching room temperature signifying the reduction in structural degrees of order suggesting a stable form at lower temperature. The particular $A_{1g}(6)$ mode, related to Cr–O stretching/bending–type mode follows only lattice anharmonicity or whether it contains an extra contribution caused by magnetic ordering (spin–phonon coupling). This mode is the most dominant, the symmetric stretching mode $A_{1g}(6)$ shows a shift of 4 cm$^{-1}$ over the whole temperature region. The next important and high intensity mode $A_{1g}(5)$ is chosen to observe the impact of temperature on mode arise due to $CrO_6$ bending coming from the rotational symmetry of [101] axis.

The temperature-dependence of the phonon-mode with frequency ω follows the conventional relation [19]:

$$\omega(T) = \omega_0(T) + \Delta\omega_{lat}(T) + \Delta\omega_{sp-lat}(T) + \Delta\omega_{el-ph}(T) + \Delta\omega_{anh}(T) \qquad 3$$

Here, the first term is the frequency of the mode at 0 K, the second terms correspond to the change in the lattice volume due to the quasi-harmonic effect, $\Delta\omega_{sp-lat}$ is related to the spin lattice coupling induced mode modification. We can omit the contribution of the fourth term $\Delta\omega_{el-ph}$ or the electron-phonon coupling to the frequency due the predictable semi conduction/insulating nature of the chromate perovskite oxides like our investigated samples. Finally, the last term is the inherent anharmonic contribution added to the phonon mode frequency which leads to the hardening of the modes below $T_N$. This equation can be overall decomposed into two terms:

$$\omega(T) = \Delta\omega_{sp-lat}(T) + \Delta\omega_{anh}(T) \qquad 4$$

as the renormalization due to magnetic ordering or spin–phonon coupling and phonon frequency shift expected from purely anharmonic phonon-phonon interactions (thermal expansion, phonon decay) are only expected possible contributions in the present samples. The increase in the wavenumber with decreasing the temperature can be explained with the help of anharmonic effect signifying the change in the intrinsic anharmonic frequency at constant volume which can be expressed as given in the Eq. (5) [17]:

$$\omega_{anh}(T) = \omega_0 + A\left[1 + \frac{2}{e^{\frac{\hbar\omega_0}{2k_BT}} - 1}\right] + B\left[1 + \frac{3}{e^{\frac{\hbar\omega_0}{3k_BT}} - 1} + \frac{3}{(e^{\frac{\hbar\omega_0}{3k_BT}} - 1)^2}\right] \qquad 5$$

with A and B are the anharmonic coefficients obtained from the Raman data fitting above $T_N$ and expression $\omega_0$ is the frequency of the corresponding mode at absolute temperature and $k_BT$ is the thermal energy.

The fitted lines are shown in the Fig. 7(c) with solid black lines. The curvatures are extrapolated below $T_N$ in order to emphasize the deviation from the fit. The vertical dotted line represents the corresponding $T_N$ and $T_{Comp}$ taken from magnetic measurements. The anomaly between the Raman phonon modes and magnetic behaviour can be due to the spin-phonon coupling or magnetostriction effect. The origin of the hardening of the particular four modes, $A_{1g}(3)$, $A_{1g}(5)$, $B_{3g}(2)$, and $A_{1g}(6)$ can be understood as a result of exchange-striction effect or the presence of $Cr^{3+}$- $Cr^{3+}$ AFM coupling evident from the shifts in wave numbers to higher values. The linewidth of these modes is related to the phonon-lifetime which stays unaltered with the volume modification of lattice due to magnetostriction but with speculative inspection leads to guess a change in slope in linewidth below the transition temperature with much fluctuation at lower temperature in both the samples. This verification confirms the existence of the strong spin-phonon coupling of $R^{3+}$-$Cr^{3+}$ and $Cr^{3+}$-$Cr^{3+}$ interaction which will later be discussed in details using the magnetic characterizations [14].

After confirming the anomalous character of the phonon modes results from the lattice contribution and/or spin-phonon coupling, one can estimate the strength of the spin-phonon coupling interaction proposed by Granado et. al. [19]:

$$\Delta \omega_{sp-lat} \propto <S_i.S_j> \qquad 6$$

Here the proportionality constant $\lambda$ is known as the spin-phonon coupling strength or coefficient and $<S_i.S_j>$ is the nearest neighbour spin-spin correlation function with $i^{th}$ and $j^{th}$ spin. And in the molecular field theory, this $<S_i.S_j>$ is quantified as follows:

$$\Delta\omega_{sp-lat} \approx \lambda S^2 \left[1-\left(\frac{T}{T_N}\right)^\gamma\right] \qquad 7(a)$$

$$\approx \lambda \left[\frac{M(T)}{M_{sat}}\right]^2 \qquad 7(b)$$

with $S$ is 3/2 (for $Cr^{3+}$), $\gamma$ is the critical exponent, M(T) is defined as temperature dependent magnetization taken from the magnetic measurements and $M_{sat}$ is the saturation magnetization for the corresponding samples. We fitted equations 6(a) and 6(b) for the particular mode $A_{1g}(6)$ which is not shown as the complex behaviour gave a floating $\lambda$ values ranging from 0.9 to 1.3 cm$^{-1}$. The complex behaviour of Raman shifts showing an abrupt decrease at initial temperatures right below $T_N$ and exhibits an increase for further lower temperatures made it difficult to extract $\lambda$ values. Right below $T_N$, when spins first order, the exchange–striction interaction destabilizes some vibrational modes, lowering their frequency indicating strong spin-phonon coupling. As long-range AFM order becomes more rigid at further low T, lattice vibrations get locked and the restoring force increases, so the mode shifts upward (hardens). Hence, the Raman mode, $A_{1g}(6)$ shows an anomalous softening just below the Néel temperature, followed by hardening at lower temperatures. This crossover reflects competition between spin–lattice coupling (reducing phonon frequency at the onset of order) and the stabilization of the exchange field at low T (increasing phonon frequency).

In order to understand the role of magnetic moments of the $R^{3+}$ atoms on the spin-phonon coupling, we have evaluated the $\mu_{eff}$ for $Sm^{3+}$ and $Ce^{3+}$ atoms to be 0.84 $\mu_B$, and 2.54 $\mu_B$ with respective $L$ values, 5, and 3 (as given in supplementary information). As given in Table 1, the $CrO_6$ octahedral distortion parameter, ($\Delta$) extracted from the refinement of the diffraction patterns of SCCO is $2.47 \times 10^{-4}$. From the $\Delta$ value of SCCO, it is understandable that the exchange integrals are more strongly modulated by oxygen displacements leading to abrupt change in wavenumber showing a clearer phonon anomaly at $T_N$ in SCCO. This could be the outcome of the exchange integral with respect to atomic displacement termed as exchange pathway (Cr–O–Cr angle and Cr–O bond lengths) which is more sensitive to the static octahedral distortion existing in the sample. Those bond-angle changes are what control the exchange interaction $J$ (Goodenough–Kanamori type dependence $J \propto t^2$ with hopping t very sensitive to bond angle and distance). Therefore, larger structural distortion in SCCO amplifies spin–phonon coupling, giving the significant observed anomaly [34, 35, 36]. The rare-earth magnetic moments and octahedral distortions jointly govern the strength of spin–phonon coupling in the present chromate perovskites. $Sm^{3+}$ with its strong SOC in SCCO is highly anisotropic, and combined with 10% of $Ce^{3+}$ leads to pronounced modulation of the Cr–O–Cr exchange pathway, and hence a stronger SPC evident from explicit deviation of low temperature Raman shifts below $T_N$.

## 4. Conclusions

In summary, $Ce^{3+}$ substitution at the Sm site in $SmCrO_3$ significantly modifies the coupled structural, magnetic, and phononic landscapes of the perovskite lattice. The increase in *A*-site ionic radius induces enhanced octahedral distortions and changes in the Cr–O–Cr exchange geometry, which in turn strengthen spin–lattice interactions. These distortions drive a pronounced reconfiguration of the magnetic ground state, enabling a field- and temperature-dependent crossover between the weak-ferromagnetic $\Gamma_4(G_x, A_y, F_z; F_z^R)$, state and the purely antiferromagnetic $\Gamma_1(A_x, G_y, C_z; C_z^R)$ configuration. The coexistence and competition of these magnetic phases result in large coercivity and a robust exchange-bias field ($H_{EB} \approx 2$ kOe), even under zero-field-cooled conditions, reflecting strong interfacial exchange between the coexisting spin manifolds. Temperature-dependent Raman spectroscopy further reveals clear phonon anomalies at $T_N$, confirming strong spin–phonon coupling mediated by $Cr^{3+}$–O vibrations and amplified by Ce-induced lattice distortions. Overall, the results demonstrate that targeted *A*-site chemical substitution provides an effective route to tune octahedral tilting, spin reorientation, magnetic anisotropy, and spin–phonon interactions in rare-earth chromites. Such tunability highlights $Sm_{0.9}Ce_{0.1}CrO_3$ as a promising platform for designing functional oxides for exchange-bias devices, magneto-thermal switches, and spintronic applications.

## 5. Acknowledgments

S. D. and R.K.D. acknowledge the FIST program of the Department of Science and Technology, India, for partial support of this work (Grant Nos. SR/FST/PSII-020/2009 and SR/FST/PSII-037/2016). S. T. acknowledges the DST SERB Core Research Grant File No. CRG/2022/006155 for the support of this work. S. T. acknowledges the Central Instrument Facility (CIF) of the Indian Institute of Technology Guwahati for partial support of this work. R.K.D. acknowledges the financial support from the Council of Scientific and Industrial Research, Ministry of Science and Technology, the Ministry of Education, Government of India. S. T. acknowledges the invités professor fellowship from the Université of Caen. S.T. thanks the North East Centre for Biological Sciences and Healthcare Engineering (NECBH) of the Indian Institute of Technology Guwahati, Assam, for partial support of this work under the grant DBT Project code: BT/NER/143/SP44675/2023.

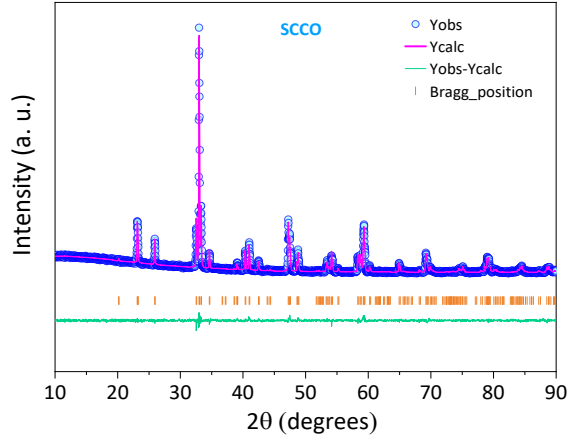

**Figure 1.** Rietveld refined X-ray diffraction patterns of SCCO showing single phase *Pbnm* orthorhombic perovskite structure.

**Table 1** The refined crystallographic parameters obtained from XRD of SCCO. The parameters, *a, b, c, and V* are lattice constants and lattice volume respectively. $\theta_1$ and $\theta_2$ are the Cr-O$_{(1)}$-Cr and Cr-O$_{(2)}$-Cr bond angles. $\theta$ and $\phi$ are the in-phase and out-of-phase tilt angles *w.r.t.* 110] and [001] respectively. $\Delta$, octahedral distortion, *t*, tolerance factor, $r_{avg}$, average radius of the $R^{3+}$ cation.

| Parameter | | Atomic positions | | Bond lengths | | Bond angles | | Other parameters | |
|---|---|---|---|---|---|---|---|---|---|
| *a* | 5.37696 | Sm/Ce (4c) | 0.99030 | Sm/Ce-O$_{(1)}$ *(two)* | 2.4701(0) | $\theta_1$ | 156.1 | $\theta$ [110] | 11.93 |
| *b* | 5.49954 | | 0.04999 | | 2.3610(0) | | | $\phi$ [001] | 11.43 |
| *c* | 7.65468 | | 0.25000 | | | $\theta_2$ | 151.6 | | |
| *V* | 226.355 | | 1.000 | Sm/Ce-O$_{(2)}$ *(six)* | 2.3694(0) *(two)* | | | $\theta$ | 6.59 |
| | | Cr (4b) | 0.00000 | | 2.7404(0) *(two)* | | | *(from lattice parameters)* | |
| | | | 0.50000 | | 2.5059(0) *(two)* | | | $\phi$ | 12.12 |
| | | | 0.00000 | | | | | *(from lattice parameters)* | |
| | | | 1.098 | | | | | | |
| | | O1 (4c) | 0.07469 | Cr-O$_{(1)}$ *(two)* | 1.9559(0) *(two)* | | | $r_{avg}$ | 1.086 |
| | | | 0.49150 | | | | | *t* | 0.8752 |
| | | | 0.25000 | | | | | | |
| | | | 0.855 | | | | | $\Delta$ (×10$^{-4}$) | 2.4743 |
| | | O2 (8b) | 0.70676 | Cr-O$_{(2)}$ *(four)* | 1.9486(0) *(two)* | | | s | 0.0225 |
| | | | 0.28226 | | 2.0178(0) *(two)* | | | $\phi$ | 13.7699 |
| | | | 0.05087 | | | | | | |
| | | | 1.008 | | | | | | |

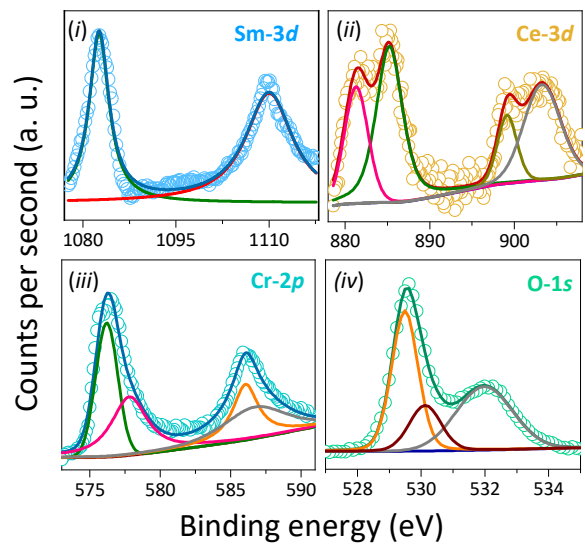

**Figure 2** X-ray photoelectron spectroscopy of SCCO (*i*) Sm−3*d*, (*ii*) Ce − 3*d*, (*iii*) Cr − 2*p* (*iv*) O − 1*s*. Scattered symbols represent the original data and solid lines are the fitted curves.

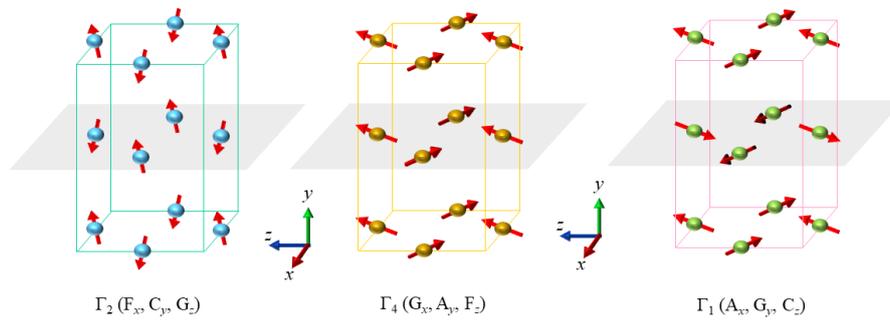

**Figure 3** Illustration of four different types of magnetic gamma spin structures discussed all through the SCCO magnetic anomalies.

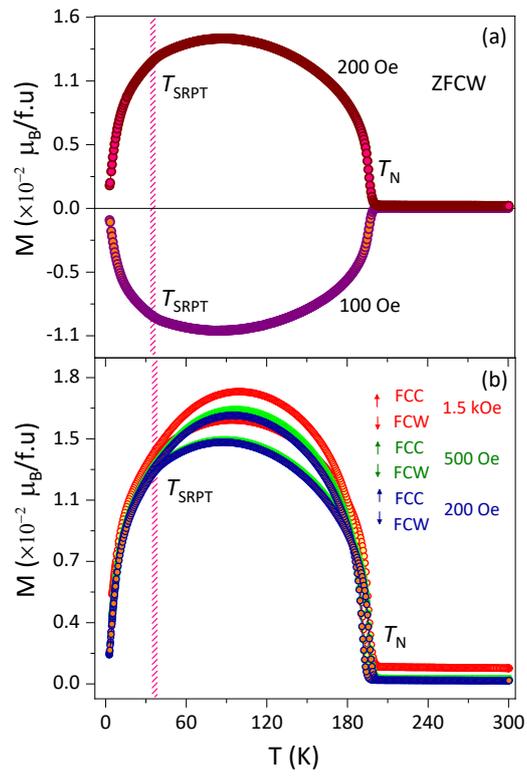

**Figure 4** Temperature dependent magnetization of SCCO under (a) ZFCW at field 100 Oe and 200 Oe. (*b*) FCC and FCW under the applied field $H$ = 200 Oe, 500 Oe, and 1500 Oe.

**Field-induced Spin-reorientation dynamics and Exchange-Bias**

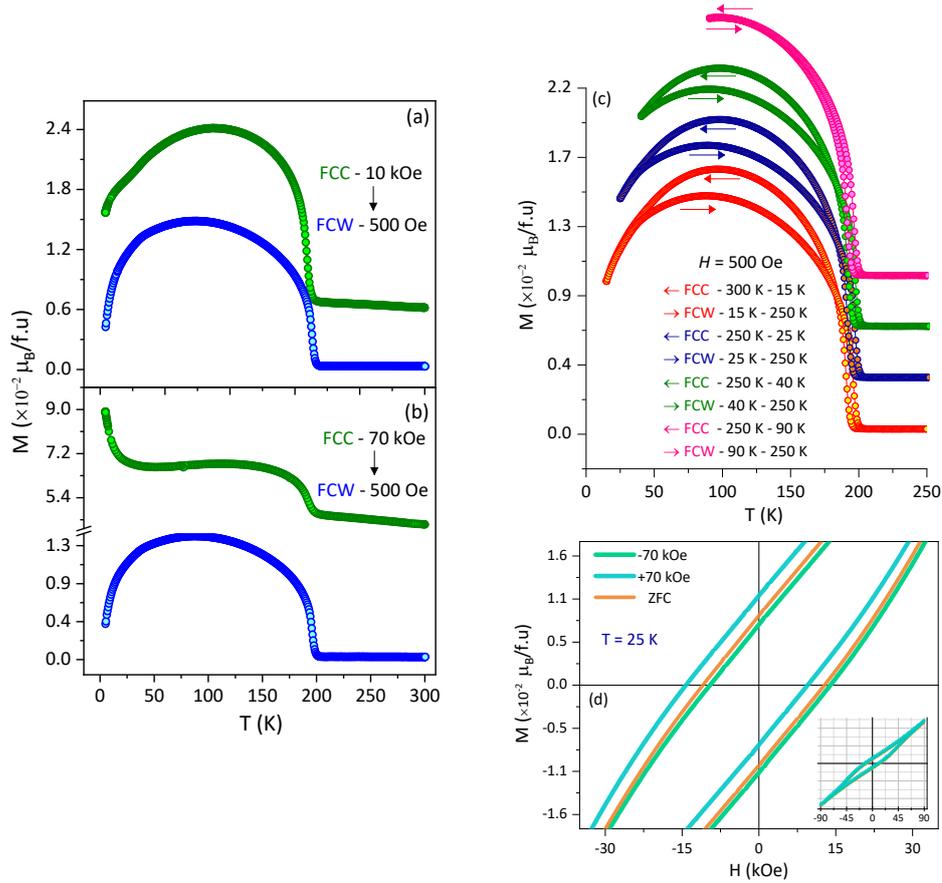

**Figure 5** (*a*) Temperature dependant magnetization under FCC at external magnetic field 10 kOe and FCW at $H_{DC}$ = 500 Oe. (*b*) Temperature dependant magnetization under FCC at $H_{DC}$ = 70 kOe and FCW at $H_{DC}$ = 500 Oe. (*c*) Kinetic arrest of Magnetic phases endearing into glassy properties. (*d*) Field dependant magnetization displaying exchange bias measured at ZFC (*Orange*), FCC at $H_{DC}$ = 70 kOe (*Purple*) and FCC at $H_{DC}$ = -70 kOe at temperature 25 K. Inset shows the MH curve at a full stretch from -90 kOe to 0 Oe to +90 kOe.

**Field-dependent Spin-reorientation from WFM-$\Gamma_4$ → Purely AFM-$\Gamma_1$**

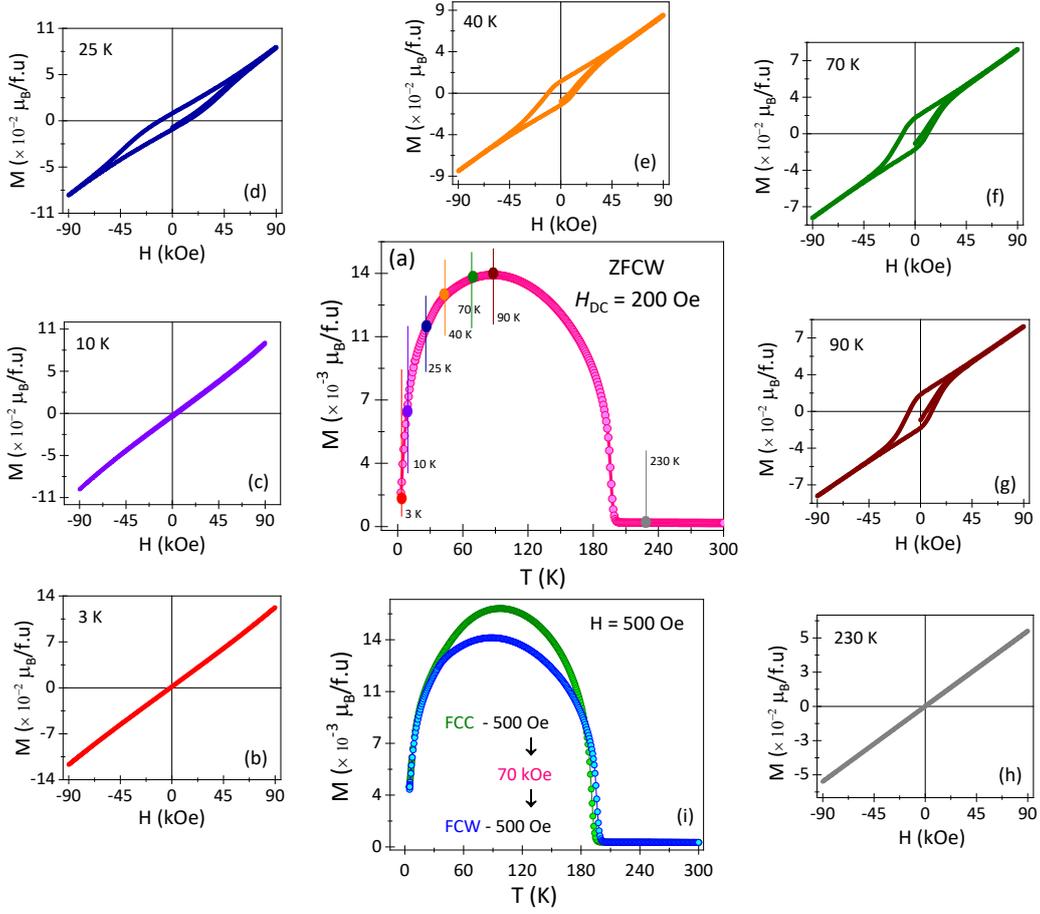

**Figure 6** (*a*) Temperature dependant magnetization in an external magnetic field $H_{DC}$ = 200 Oe under ZFCW. Field dependent magnetization of SCCO at (*b*) 3 K, (*c*) 10 K, (*d*) 25 K, (*e*) 40 K, (*f*) 70 K, (*g*) 90 K, (*h*) 230 K. (*i*) Temperature dependent magnetization under FCC at $H_{DC}$ = 500 Oe, at 3 K, the field is increased to 70 kOe and decreased to 500 Oe and FCW is measured at $H_{DC}$ = 500 Oe.

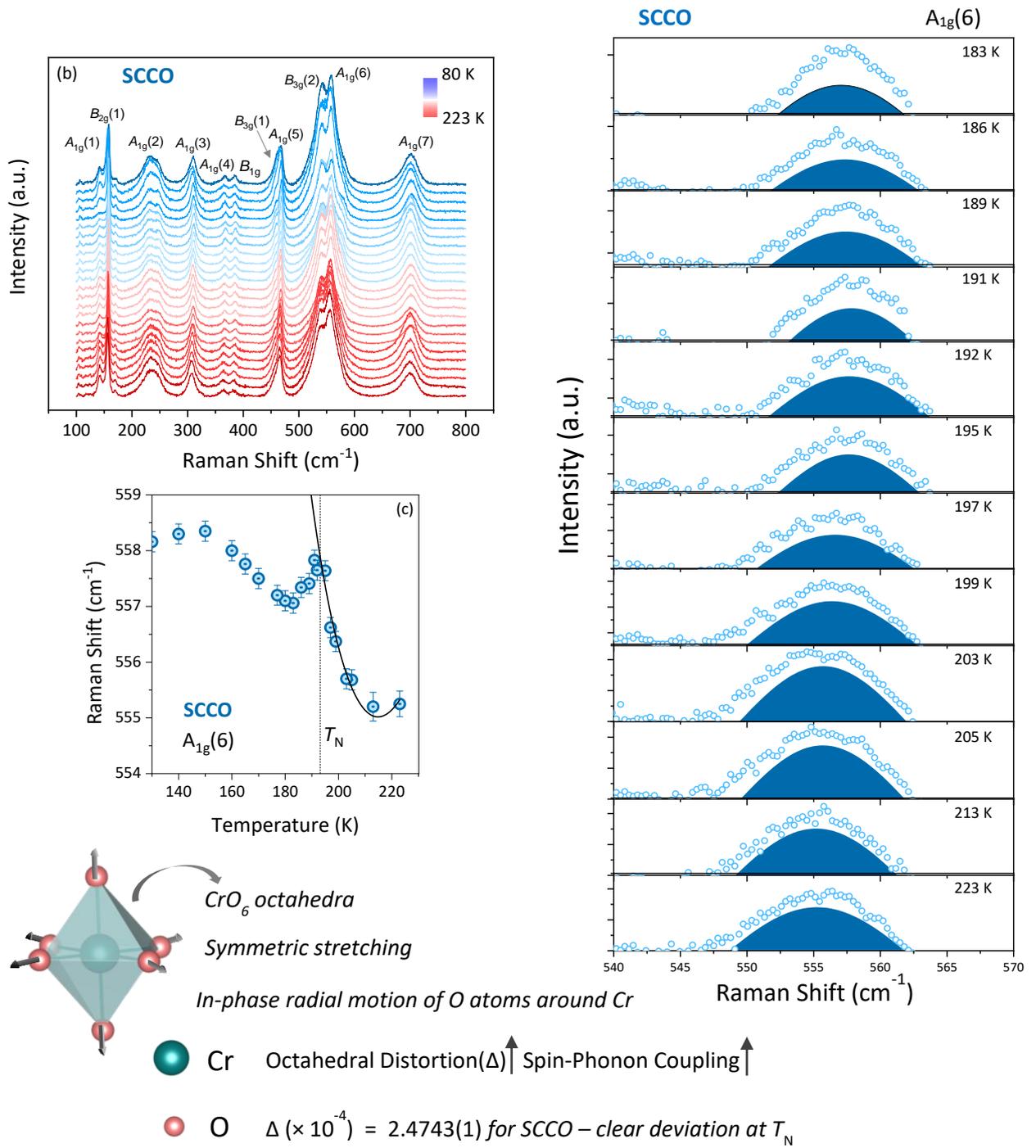

**Figure 7** (a) Raman spectra with all the existing modes indexed for SCCO from 80 K to 223 K. Self-explanatory schematic of symmetric stretching mode, A$_{1g}$(6) response to temperature in and corresponding Lorentzian fits.

# Supplementary Information

# Spin-Reorientation Dynamics and Strong-Spin Phonon Coupling in Ce-substituted SmCrO$_3$

Shaona Das[1,3*†], Ravi Kiran Dokala[2,3*†] and Subhash Thota[3*]

[1]Institute of Material Science, Technical University of Darmstadt, Darmstadt-64287, Germany
[2]Solid State Division, Department of Materials Science and Engineering, Uppsala University, Uppsala-75237, Sweden
[3]Department of Physics, Indian Institute of Technology Guwahati, Guwahati, Assam-781039, India

Supporting information to the presence of Spin-Phonon Coupling, SPC:

Effective magnetic moments estimated from Hund's rules for the rare earth elements, Sm$^{3+}$, and Ce$^{3+}$ in SCCO:

Sm$^{3+}$ ($4f^5$, $L = 5$, $S = 5/2$, implies $J = |L - S| = 5/2$, $g_J = 2/7$ with spin-only ground state $H^6_{5/2}$):

$$\mu_{\text{eff}} = g_J\sqrt{J(J+1)}$$
$$\mu_{\text{eff}} \approx 0.84\ \mu_B$$

Ce$^{3+}$ ($4f^1$, $L = 3$, $S = 1/2$, implies $J = 5/2$, $g_J = 6/7$ with ground state $F^2_{5/2}$):

$$\mu_{\text{eff}} \approx 2.54\ \mu_B$$

In octahedral field CrO$_6$, Cr$^{3+}$ ($3d^3$, $L_{\text{eff}} = 0$, $S = 3/2$ with $t_{2g}^3 e_g^0$, giving $J = 3/2$) :

$$\mu_{\text{eff}} \approx 3.87\ \mu_B$$

SPC originates from the phonon vibrations modulate the magnetic exchange interaction (J$_{ex}$) between the neighbouring spins

$$\Delta\omega_{sp-lat} \propto \lambda <S_i.S_j>$$

$$\lambda \sim \frac{\partial J_{ex}}{\partial Q}$$

where Q is the phonon coordinate (for eg. bond length/bond angle, etc.).